\documentclass{PoS}

\usepackage{floatrow}

\newfloatcommand{capbtabbox}{table}[][\linewidth]

\title{Disconnected diagrams with twisted-mass fermions}

\ShortTitle{Disconnected diagrams tmQCD}

\author{Abdou Abdel-Rehim$^a$, Constantia Alexandrou$^{ab}$,Martha Constantinou$^{ab}$\footnote{Current affiliation: Temple University}, Jacob Finkenrath$^{a}$, Kyriakos Hadjiyiannakou$^{c}$, Karl Jansen$^d$, Christos Kallidonis$^a$, Giannis Koutsou$^a$, \speaker{Alejandro Vaquero Avil\'es-Casco}$^e$\footnote{Current affilitaion: Department of Physics and Astronomy, University of Utah}, Julia Volmer$^{d}$\\
     \\
     \llap{$^a$}Computation-based Science and Technology Research Center, CaSToRC\\
     The Cyprus Institute, 20 Kavafi Str. Nicosia 2121, Cyprus\\
     \\
     \llap{$^b$}Department of Physics, University of Cyprus\\
     P.O. Box 20537, 1678 Nicosia, Cyprus\\
     \\
     \llap{$^c$} The George Washington University\\
     2121 I St NW, Washington, DC 20052, USA\\
     \\
     \llap{$^d$}NIC, DESY\\
     Platanenallee 6, D-15738 Zeuthen, Germany\\
     \\
     \llap{$^e$}INFN Sezione di Milano-Bicocca\\
     Edificio U2, Piazza della Scienza 3, 20126 Milano, Italy\\
     \\
     E-mail: \email{a.abdel-Rehim@cyi.ac.cy}, \email{alexand@ucy.ac.cy}, \email{marthac@temple.edu}, \email{j.finkenrath@cyi.ac.cy}, \email{hadjiyiannakou.kyriakos@ucy.ac.cy}, \email{karl.jansen@desy.de}, \email{c.kallidonis@cyi.ac.cy}, \email{g.koutsou@cyi.ac.cy}, \email{alexvaq@physics.utah.edu}, \email{Julia.Volmer@desy.de}}

\abstract{The latest results from the Twisted-Mass collaboration on disconnected diagrams at the physical value of the pion mass are presented. \
In particular, we focus on the sigma terms, the axial charges and the momentum fraction, all of them for the nucleon. A detailed error analysis \
for each observable follows, showing the strengths and weaknesses of the one-end trick. Alternatives are discussed.}

\FullConference{34th annual International Symposium on Lattice Field Theory\\
		24-30 July 2016\\
		University of Southampton, UK}

\begin{document}

\section{Introduction}

During the last 10 years there has been a tremendous progress in the computation of the disconnected contributions entering the evaluation of nucleon
matrix elements. The computation of the all-to-all propagator is the major difficulty in this calculation, but new methods and powerful computers are
enabling the computation of fermionic disconnected loops to sufficient accuracy. 

In this work we show the final results on the direct evaluation of disconnected loops on a $N_f=2$ ensemble at the physical value of the
pion mass. 

\section{Noise reduction techniques}
\label{sec:disM}
%
%

The standard approach to compute disconected diagrams involves a stochastic estimation of the all-to-all propagator, generating and inverting random source
vectors~\cite{stoch,NoiseZN}. Usually this method converges poorly, with errors behaving as $O(1/\sqrt{N_r})$, where $N_r$ is the number of stochastic
sources employed, so a large $N_r$ is required. But each stochastic source must be inverted through the propagator, making the method extremely expensive
in some cases.

\subsection{The Truncated Solver Method (TSM)\label{subs:tsm}}


The TSM method~\cite{TSM} allows to circumvent this problem using a predictor-corrector scheme. First we compute a sloppy prediction (low precision LP,
as opposed to high precision HP) for the estimate stochastically, and to make it sloppy we invert the source to very low precision. Usually this is
achieved by stopping the inverter (a CG in our case) at a fixed value of the residual $\rho_{LP} \gg \rho_{HP}$ or of the number of iterations
$n_{LP} \ll n_{HP}$, that is far from convergence. Hence the inversions become cheap and one can increase statistics without incurring in large computer
costs.

In order to correct the bias introduced by our sloppy inversions, another set of stochastic sources is
generated. Each one of these sources are inverted to both sloppy and high precision, and the difference is calculated. The average over this
set of sources gives the correction,
\begin{equation}
G_{E_{TSM}}:=\underbrace{\frac{1}{N_{\rm HP}}\sum_{j=1}^{N_{\rm HP}}
\left[\left|s_j\right\rangle_{HP} - \left|s_j\right\rangle_{LP}\right]
\left\langle\eta_j\right|}_{\textrm{Correction}} + \underbrace{\frac{1}{N_{\rm LP}}
\sum_{j=N_{\rm HP}+1}^{N_{\rm HP}+N_{\rm LP}}\left|s_j\right\rangle_{LP}
\left\langle\eta_j\right|}_{\textrm{Biased estimate}},
\label{estiTSM}
\end{equation}
Since the correction involves high precision inversions, it is expensive. Nonetheless, if the sloppy and the high precision solutions for the same noise
source are closely correlated, only a few rhs are needed to calculate an accurate correction. Then the bulk of the computation goes to the
sloppy estimate, and the errors decrease approximately as $O(1/\sqrt(N_{LP}))$.

As the quark mass decreases, the number of iterations required to achieve a large enough correlation increases. This fact affects negatively the
performance of the TSM.

\subsection{Deflation\label{subs:def}}

Deflation can help overcome the decrease of performance of the TSM for light quark masses by removing the low-modes from the computation: first we
calculate the $N_{ev}$ lowest eigenpairs
$\lambda_j, \left|v_j\right\rangle$ of $M$. The low-modes generate a subspace where the original noise source can be projected, 
removing the low modes from the stochastic source, and thus leading to a faster inversions due to the improvement
conditioning of the matrix. Once inverted, the low-mode part of the solution can be exactly reconstructed from the eigenpairs.


\subsection{Low-mode reconstruction\label{subs:low}}

\begin{figure*}[h!]
\begin{floatrow}
\ffigbox{
  \vspace{-0.25cm}
  \includegraphics[width=\linewidth,angle=0]{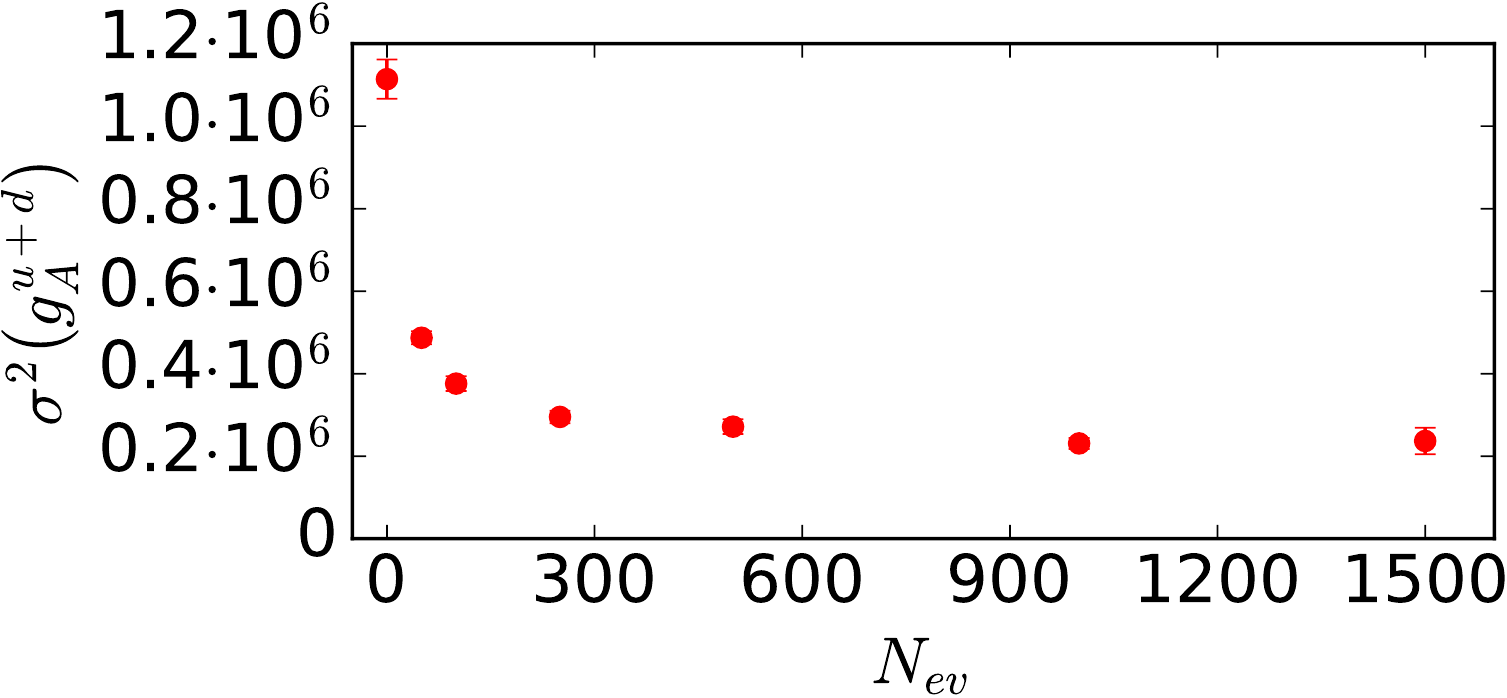}
}
{
  \vspace{-0.65cm}
  \caption{\footnotesize Evolution of the variance for the nucleon $g_A^{u+d}$ as the number of low modes for the exact part increases. There is a
           saturation from $\approx 250$ modes on, where the rest of the modes contribute with similar weights, and there are no further gains. In our
           tests the most efficient approach didn't reach the saturation point.\label{figDef}}
}
\capbtabbox{
  \vspace{-0.25cm}
  \begin{tabular}{l|c|c}
    Method						   & $N_{st}$ & Cost \\
    \hline
    {\color{blue} $N_{ev}^{EO}=500$}         		   &   2250   & 1.00 \\
    $N_{ev}^{Full}=100$ (LM)               		   &    750   & 1.54 \\
    $N_{ev}^{Full}=250$ (LM)               		   &    600   & 0.97 \\
    {$\color{red}N_{ev}^{EO}=500, N_{ev}^{Full}=100$ (LM)} &    750   & 0.61 \\
    $N_{ev}^{EO}=500, N_{ev}^{Full}=250$ (LM)              &    600   & 0.77 \\
  \end{tabular}
  \vspace{1.0cm}
}{
  \vspace{-0.65cm}
  \caption{\footnotesize Cost and number of rhs required to reach a fixed error in the different deflation schemes, where LM stands for
           \emph{Low Mode reconstruction}. The cost is normalized to that of the blue method, which is the one we used for the ultralocal
           quantities at light masses. The one-derivative operators used the more efficient red method.\label{defTb}}
}
\end{floatrow}
\end{figure*}

The previous method can be further improved if instead of reconstructing $\left|s^L\right\rangle$ the all-to-all propagator corresponding to the low-modes
is computed~\cite{Klaus}. In this method only $\left|s^D\right\rangle$ is calculated, contracted and averaged. Then the exact low-mode part of the
all-to-all can be reconstructed.
This method removes all the stochastic noise coming from the low-modes, but it requires knowledge of the eigenvectors of the full operator, as opposed to
the Even-Odd (EO) precondiitoned one.
In order to choose the best approach, we tried several combinations summarized in Table~\ref{defTb}. In our experience, it was more advantageous to compute
the eigenvalues of both operators (full and preconditioned), to reduce the variance of the loops and the cost of the inversions.

\subsection{The one-end trick\label{subs:vvt}}

The one-end trick \cite{vvTrick1,vvTrick2} for disconnected diagrams is a variance reduction technique that can be applied in twisted-mass fermions
to reduce the stochastic errors at no extra cost. In the twisted basis, isovector operators can be casted as a product of propagators,
\begin{equation}
\Gamma\left[G_u - G_d\right] = -2i\mu a\Gamma G_d\gamma_5G_u.
\end{equation}
A similar construction can be found for isoscalar quantities in the twisted-basis.
\begin{figure*}[h!]
   \begin{center}
      \begin{minipage}{0.4\linewidth}
        \includegraphics[width=\linewidth,angle=0]{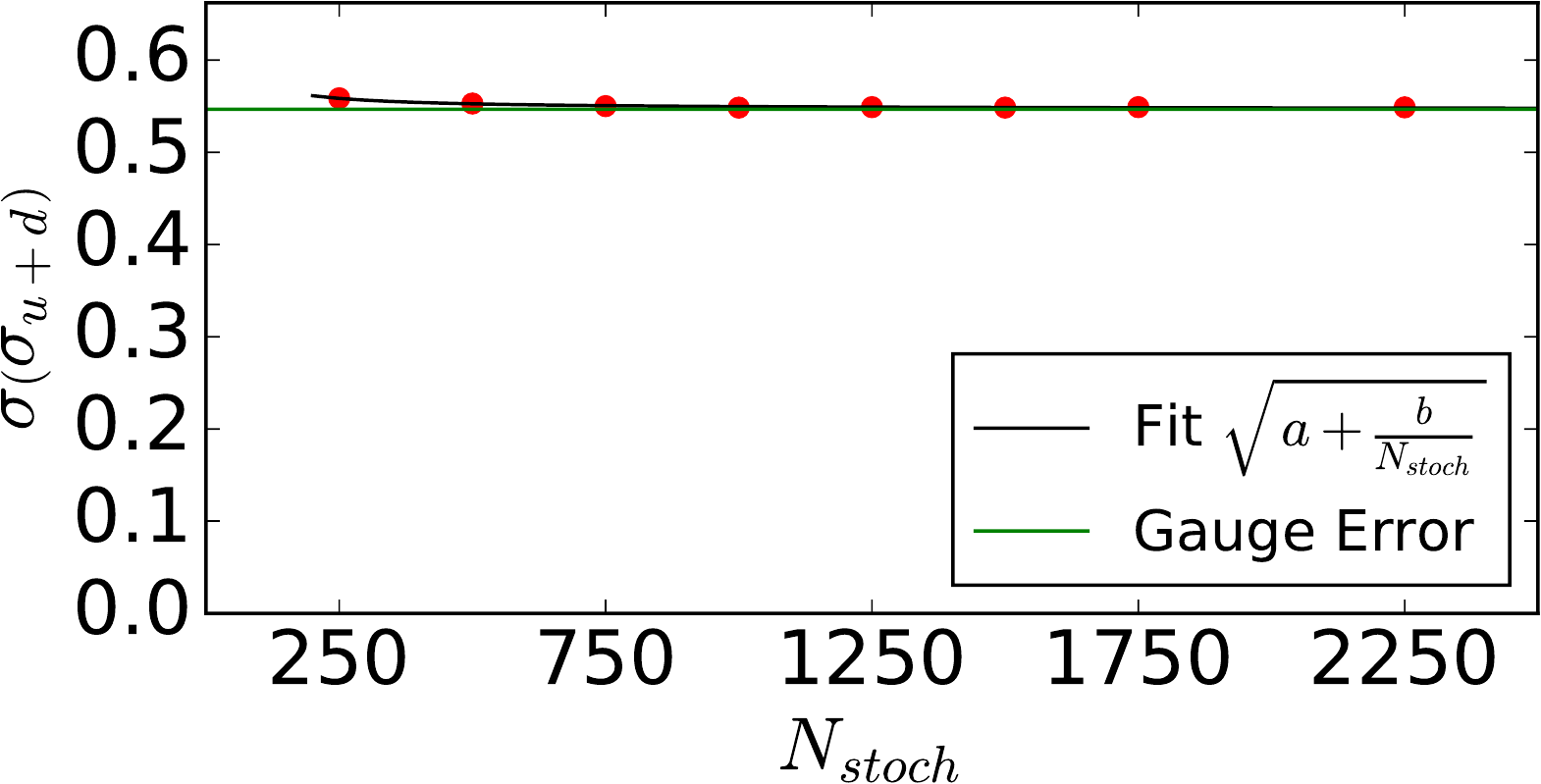}
        \includegraphics[width=\linewidth,angle=0]{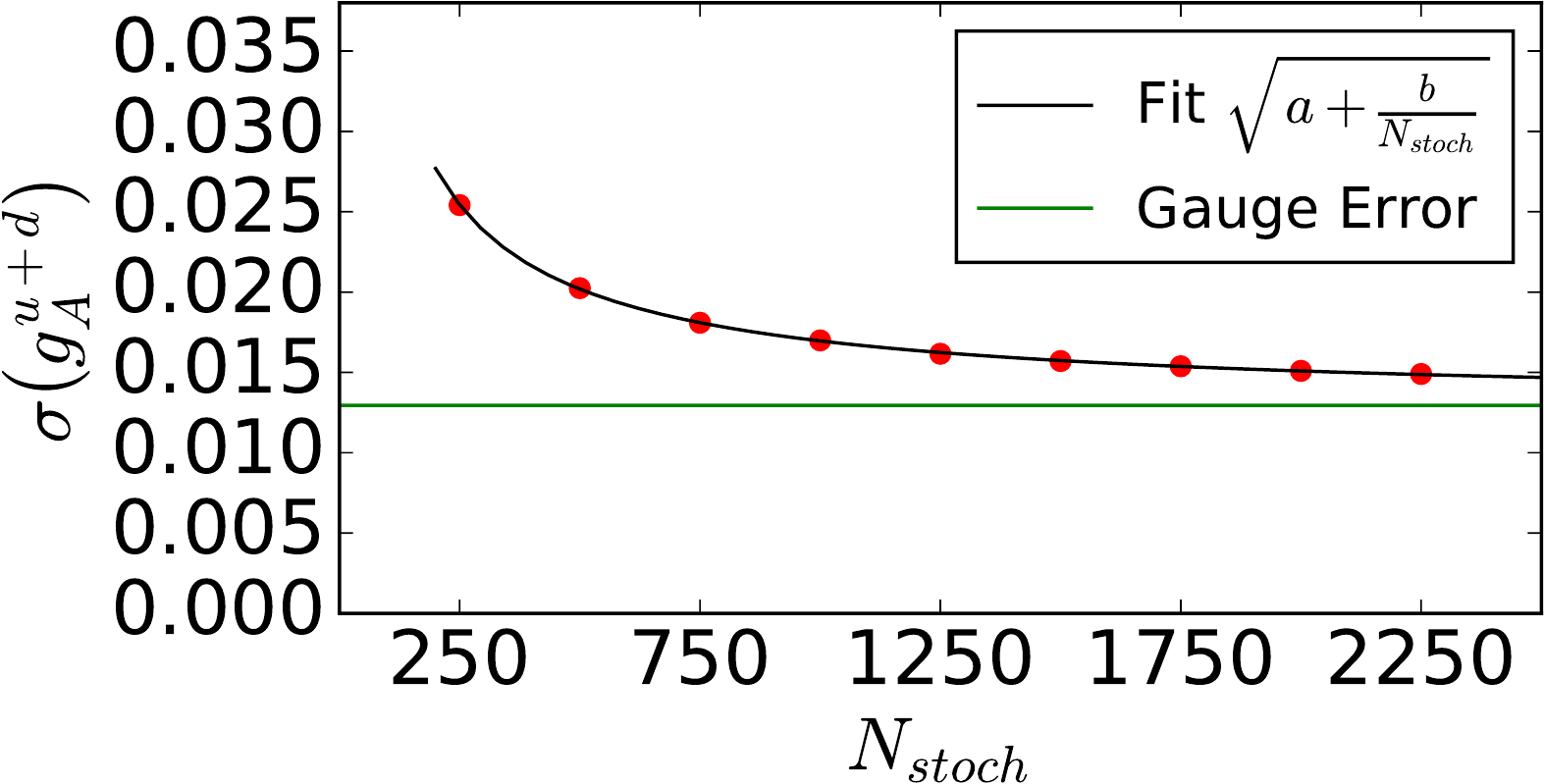}
      \end{minipage}
      \hspace{0.05\linewidth}
      \begin{minipage}{0.4\linewidth}
        \includegraphics[width=\linewidth,angle=0]{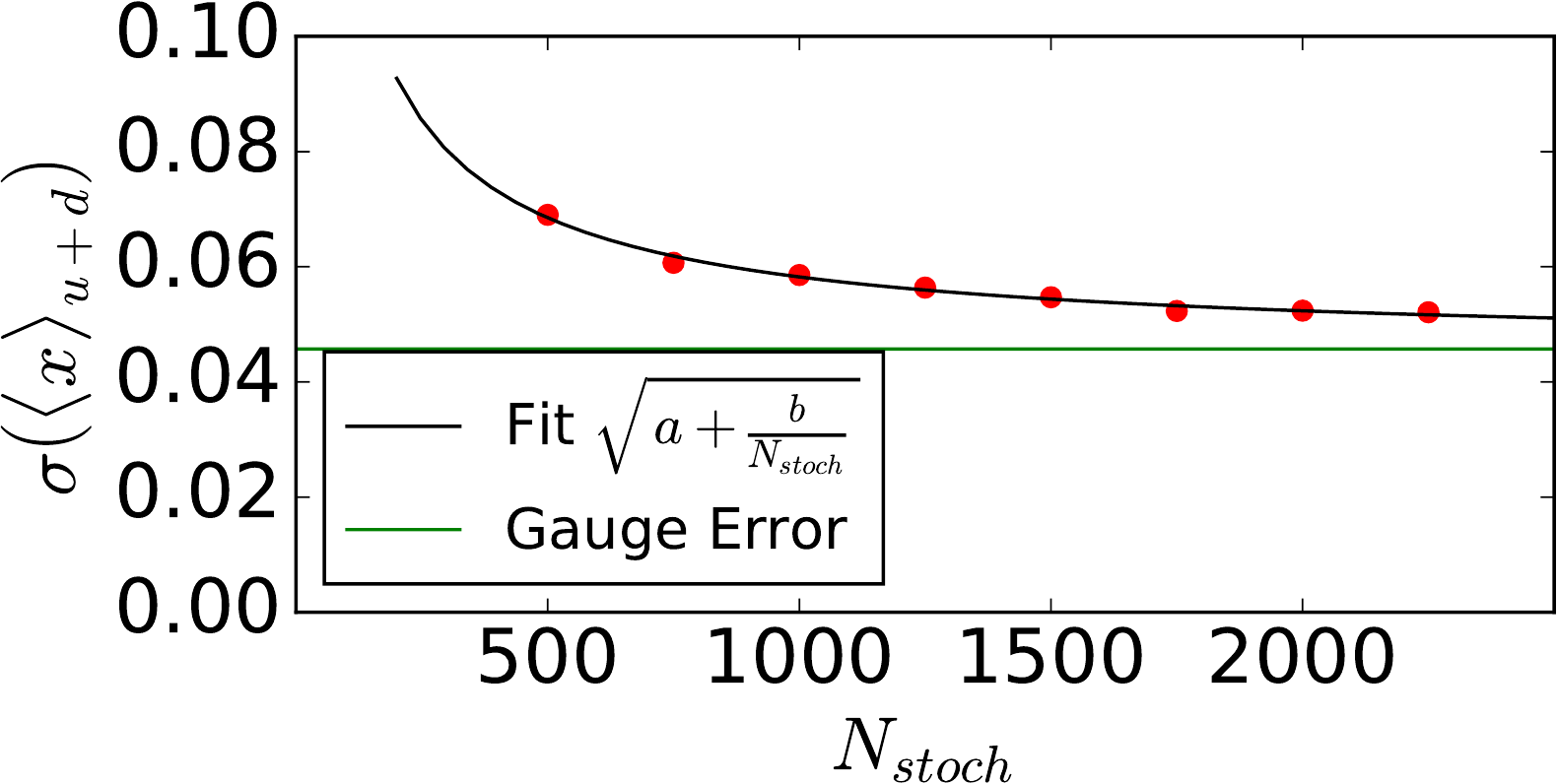}
        \includegraphics[width=\linewidth,angle=0]{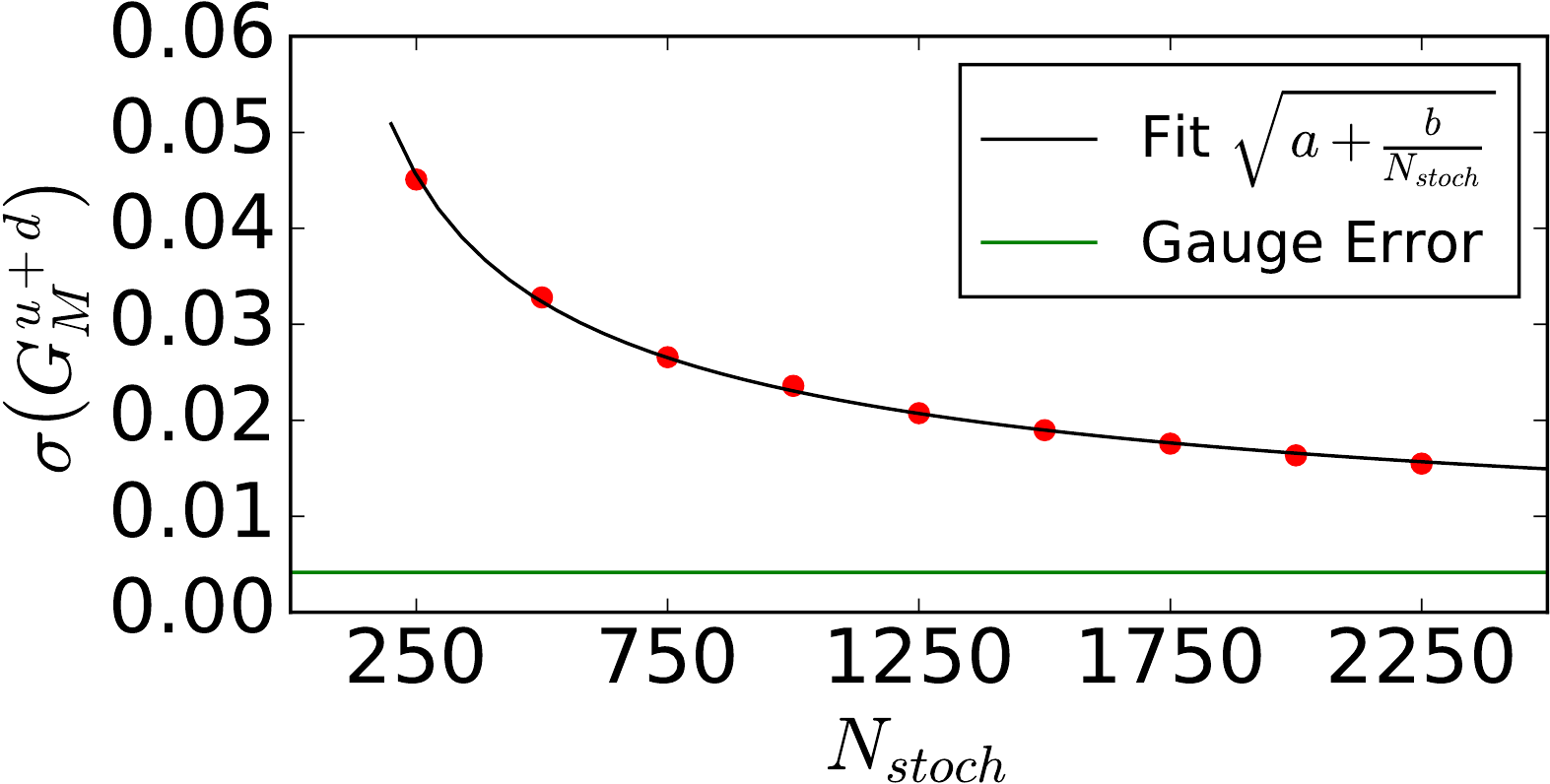}
      \end{minipage}
      \vspace{-0.5cm}
      \caption{\footnotesize Total absolute error for the disconnected parts of $\sigma_{u+d}$, $g_A^{u+d}$, $\left\langle x\right\rangle_{u+d}$, and
	       $G_M^{u+d}$, as a function of the number of stochastic sources.\label{err}}
   \end{center}
\end{figure*}
In fig.\ref{err} we assess the performance of the one-end trick for a few observables, with no other variance reduction technique. Of those, only the
$\sigma-$term uses the more powerful, isovector version of the trick. As we see, the $\sigma-$term shows a superb performance, and we could cut the cost
of the computation by a factor of 10 and still obtain virtually the same errors. The other observables use the less powerful isoscalar version of the
trick, which allows us to keep stochastic errors well under control in most cases, being the electromagnetic form factors a notable exception.

\subsection{Hierarchical probing\label{subs:hpr}}

As evidenced by fig.\ref{err}, the electromagnetic form factors require a different treatment.
Probing and hierarchical probing techniques \cite{Pena,meinel} have been shown to reduce very effectively the stochastic noise of the electromagnetic
form factor.

Since we wanted to keep the one-end trick and to compute the loop in all time-slices for the analysis,
we decided to try a 4D hierarchical probing with the one-end trick (in contrast to what is done in \cite{meinel}) against
our current technique (TSM + one-end trick) in an ensemble with a larger pion mass ($m_\pi\approx380$ MeV). We also assessed the effect of spin+color
dilution, which is not currently used in our production runs. The results for the nucleon axial charge $g^{u+d}_A$ and the magnetic form factor $G^{u+d}_M$
are shown in table~\ref{hProb}.

\begin{table}[h!]
  \centering
  \begin{tabular}{l|c|c}
    Method & $E_{g_A}$ & $E_{G_M}$ \\
    \hline
    Simple stochastic               & $3.073 \pm 0.012$ & $8.85 \pm 0.11$\\
    Hierarchical probing            & $2.24  \pm 0.11$  & $114  \pm 2$   \\
    Hierarchical probing + dilution & $1.35  \pm 0.05$  & $3.0  \pm 0.4$ \\
    TSM                             & $0.63  \pm 0.07$  & $2.45 \pm 0.04$
  \end{tabular}
  \caption{\footnotesize Comparison of $E=\sigma^2\times Cost$ for different methods. Lower values are more efficient.}
  \label{hProb}
\end{table}
It is notable, in our particular version of this technique, the regression of hierarchical probing without dilution for the magnetic form factor. 
As a conclusion, we don't see any
special advantage with respecto to the TSM, hence for this run we decided to stay with our current methods and ommit the electromagnetic form factors from
the calculation.

\section{Details of the simulation and the analysis}
\label{sec:setUp}
We used a $48^3\times 96$ ensemble of $N_f=2$ twisted-clover fermions, tuned at the physical value of the light quark mass ($m_\pi\approx 131$ MeV~
\cite{ETMC}), with $a = 0.093$ fm. Each contribution was combined with 100 nucleon 2-point functions per configuration. Averaging over forward-backward
and proton-neutron resulted in 400 independent measurements per configuration.

The low modes were computed using PARPACK binded to QUDA. Table~\ref{defTb} shows the number of low modes calculated. The stochastic part of the loops as
well as the contractions were always computed using QUDA \cite{QUDA,Yo2}.

We used three different analysis methods to remove contamination from the excited states: $(1)$ a plateau fit at fixed sink time, $(2)$ the summation
method for several fit ranges and $(3)$ a two-state fit for several values of the sink. In the two-state fit, we would use the same parameters to fit all
the plateaux up to the chosen sink. Once the three different estimates were obtained, we tried to find agreement between all the methods, as well as
convergence of the final value as we changed the fit parameters and ranges in the three methods.

\begin{figure*}[h!]
   \begin{center}
      \begin{minipage}{0.4\linewidth}
        \includegraphics[width=\linewidth,angle=0]{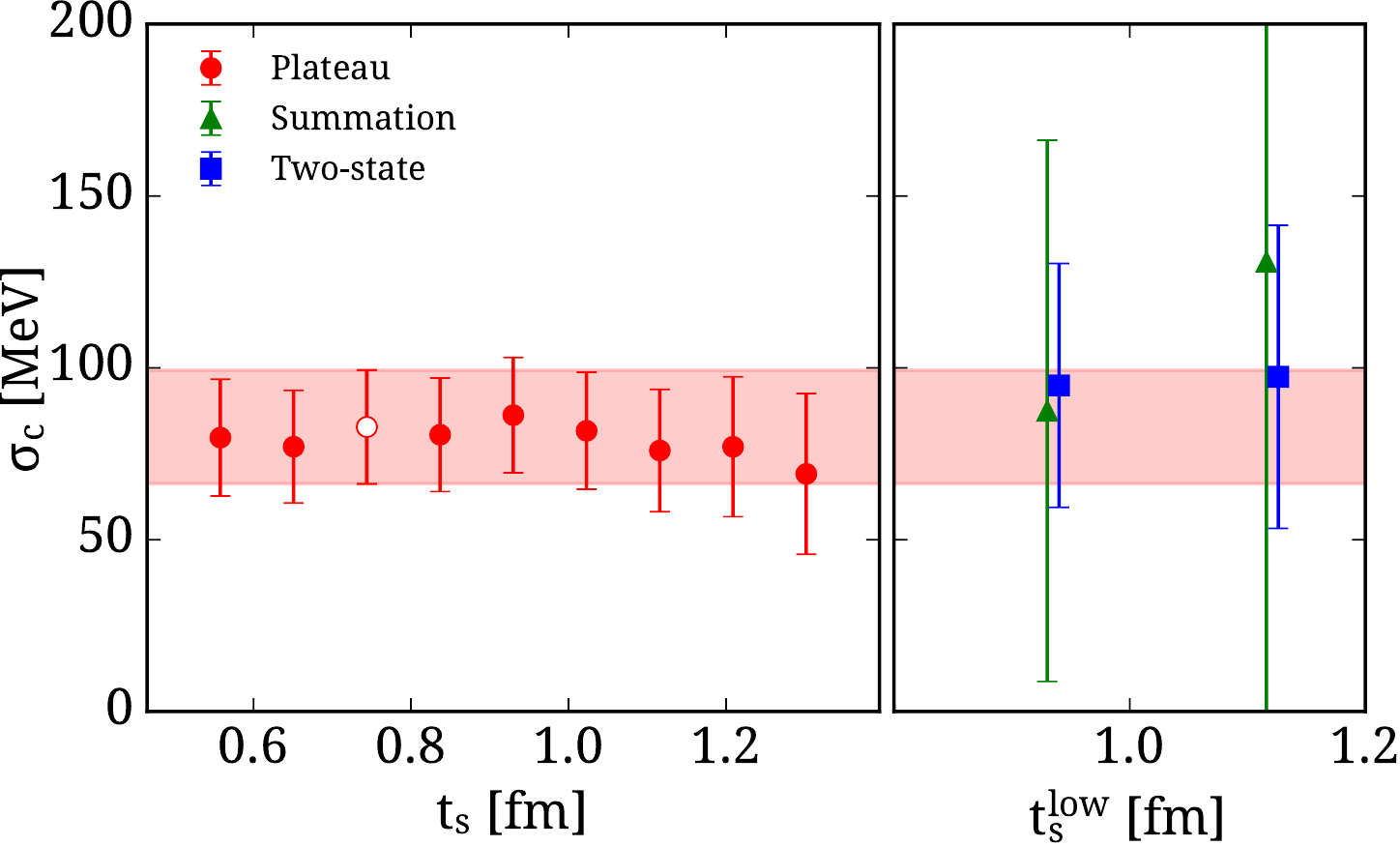}
      \end{minipage}
      \hspace{0.05\linewidth}
      \begin{minipage}{0.4\linewidth}
        \includegraphics[width=\linewidth,angle=0]{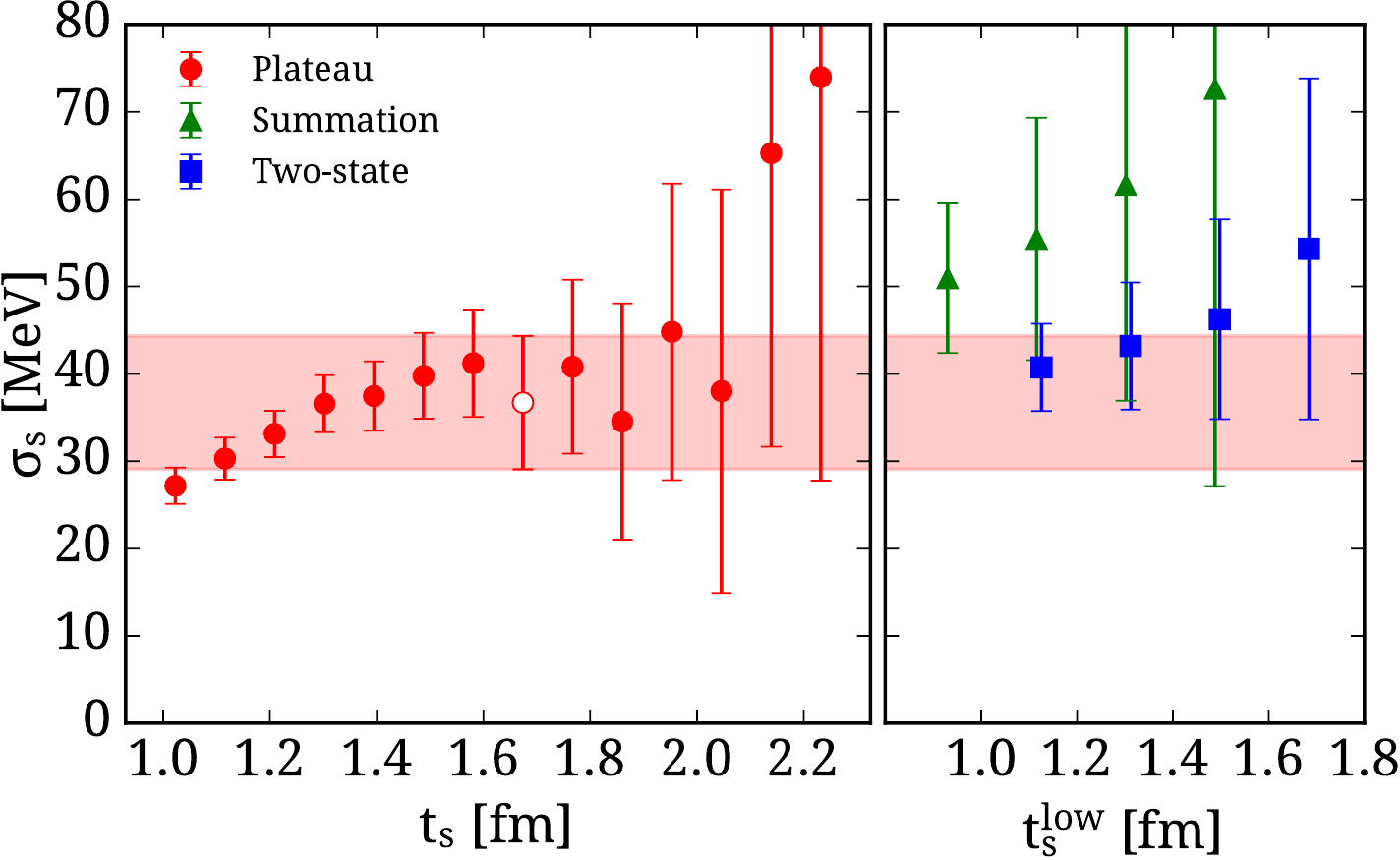}
      \end{minipage}
      \vspace{-0.5cm}
      \caption{\footnotesize Example of analysis to remove the excited states, with $\sigma_c$ on the left and $\sigma_s$ on the right. $\sigma_c$ is a
	       paradigm of well-behaved quantity, where the three methods agree on a value and converge fast. The case of, $\sigma_s$ in contrast, shows
	       a much stronger contribution of excited states, resulting in larger errors in the determination of the final value.\label{anaPlot}}
   \end{center}
\end{figure*}

\section{Results}
\label{sec:res}

For each value of the quark mass, we calculated the nucleon $\sigma-$term, axial and tensor charges, and for the light and the strange we also calculated
the momentum fraction. In general, we observe large contributions from the excited states coming from the $\sigma-$terms, being $\sigma_c$ a notable
exception. Other quantities are less prone to these controibutions, but show larger statistical errors. 

\begin{figure*}[h!]
   \begin{center}
      \begin{minipage}{0.4\linewidth}
        \includegraphics[width=\linewidth,angle=0]{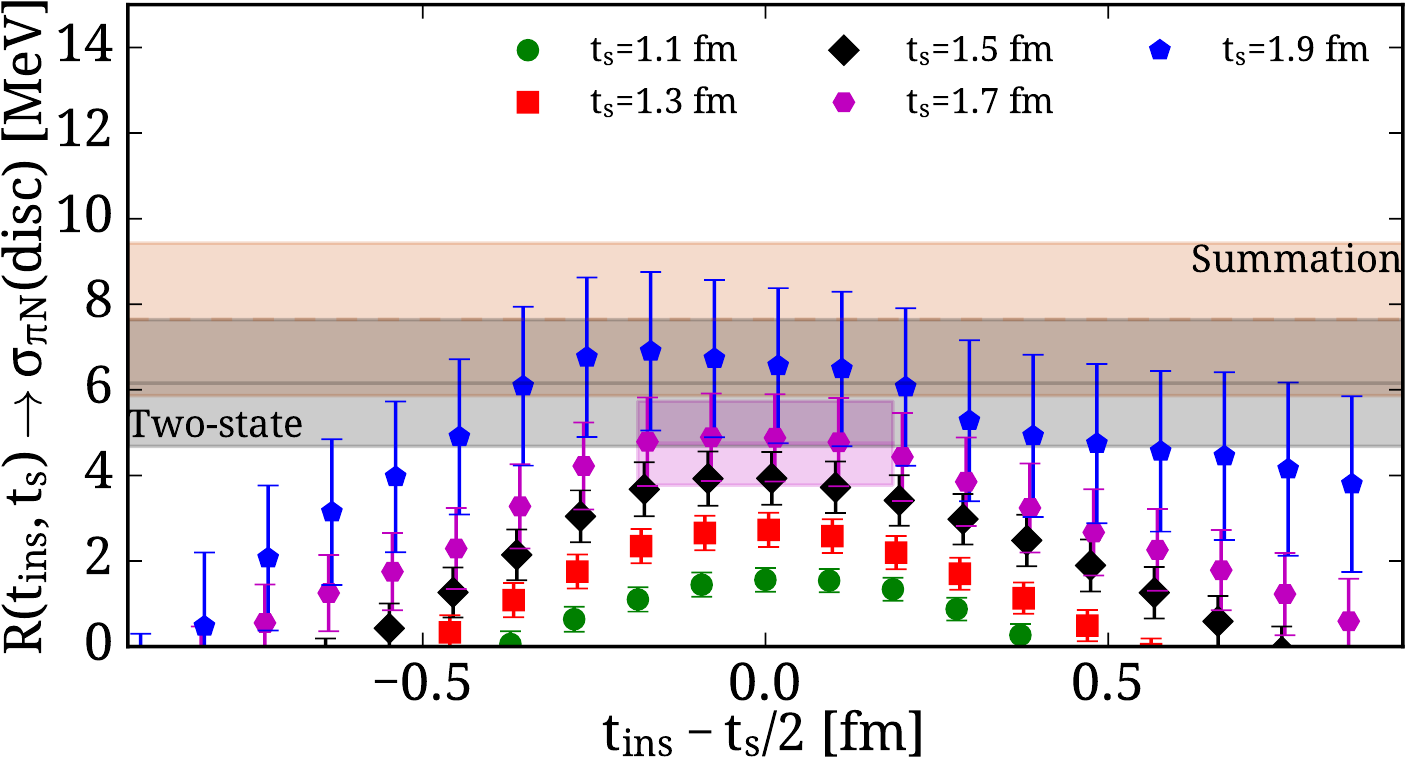}
        \includegraphics[width=\linewidth,angle=0]{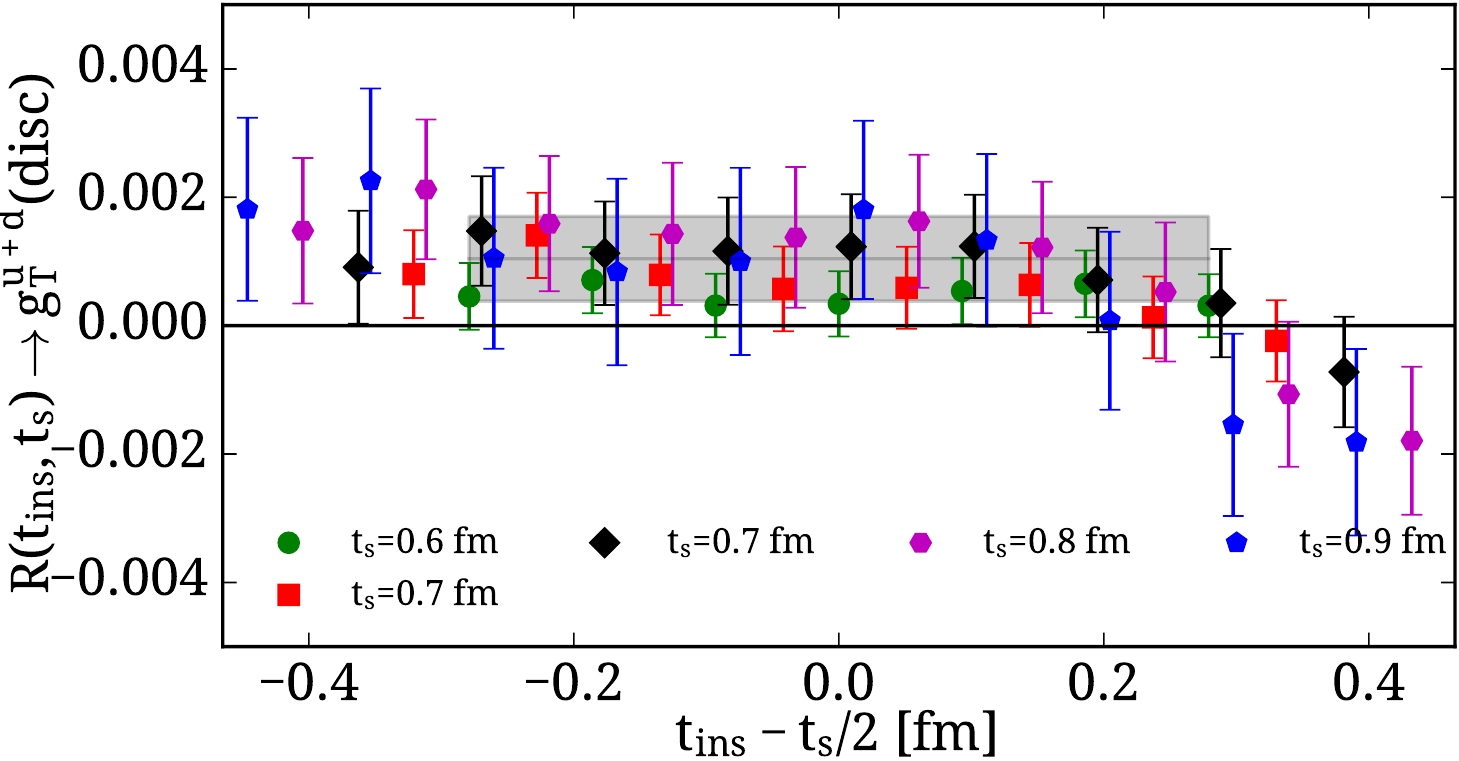}
      \end{minipage}
      \hspace{0.05\linewidth}
      \begin{minipage}{0.4\linewidth}
        \includegraphics[width=\linewidth,angle=0]{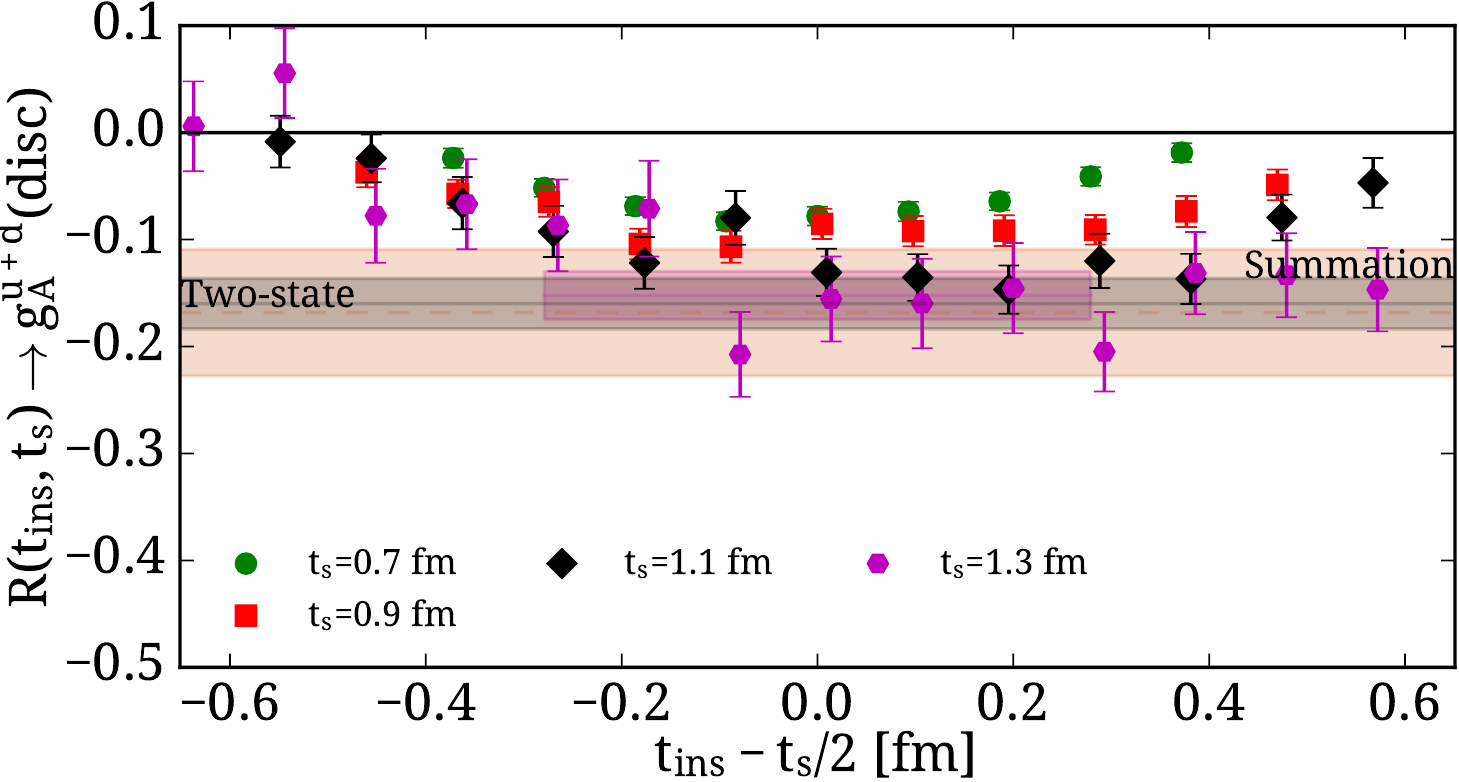}
        \includegraphics[width=\linewidth,angle=0]{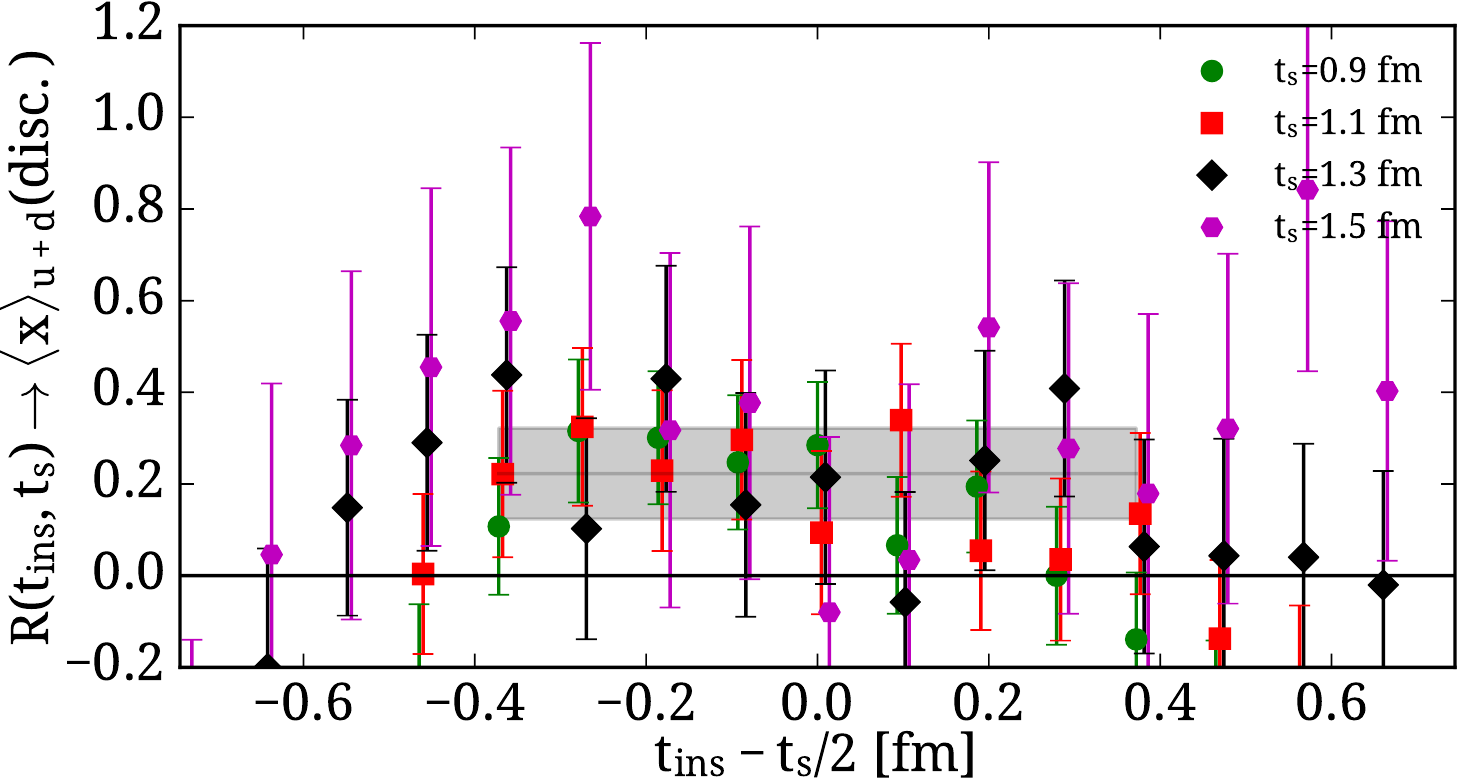}
      \end{minipage}
      \vspace{-0.5cm}
      \caption{\footnotesize Disconnected quantities for the light quark mass. For the ultralocal we used 2250 stochastic sources accelerated with a
               deflated solver over 2136 configurations, giving a total of 854400 measurements. The one-derivative results were computed in a new run with
               low mode reconstruction with 100 eigenvectors, and 1000 stochastic sources using a deflated solver for the high modes, over 1219
               configurations, resulting in 487600 measurements.\label{plotLgt}}
   \end{center}
\end{figure*}

\begin{figure*}[h!]
   \begin{center}
      \begin{minipage}{0.4\linewidth}
        \includegraphics[width=\linewidth,angle=0]{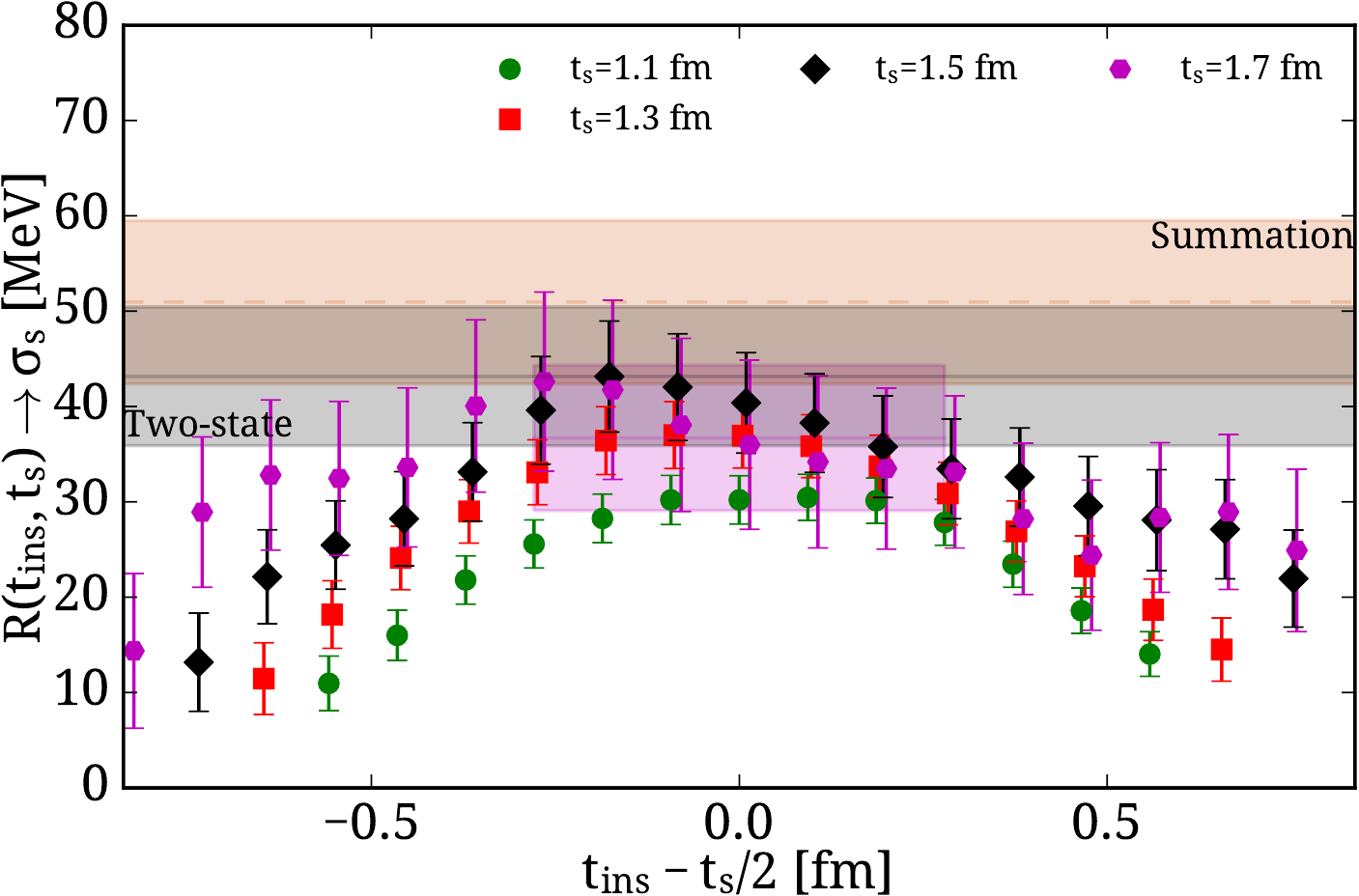}
        \includegraphics[width=\linewidth,angle=0]{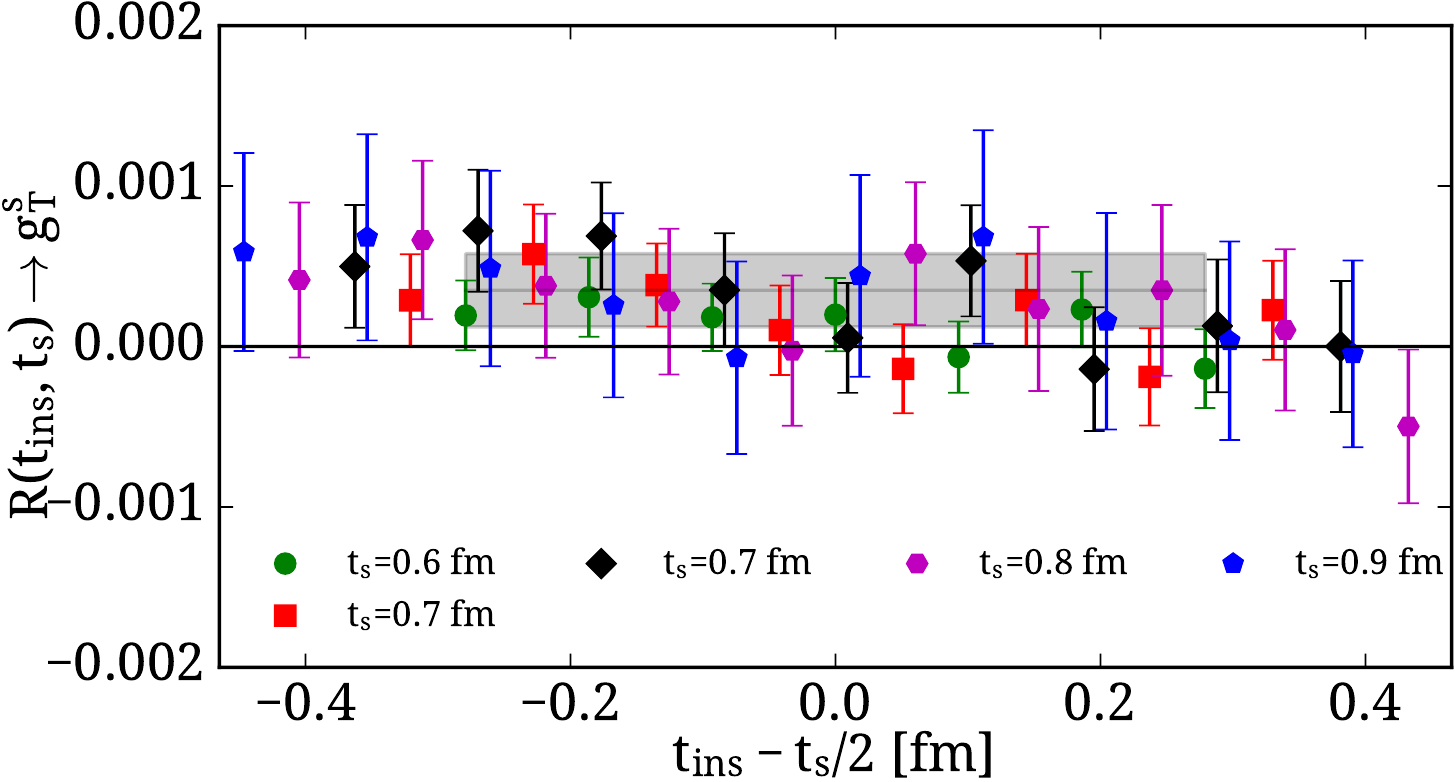}
      \end{minipage}
      \hspace{0.05\linewidth}
      \begin{minipage}{0.4\linewidth}
        \includegraphics[width=\linewidth,angle=0]{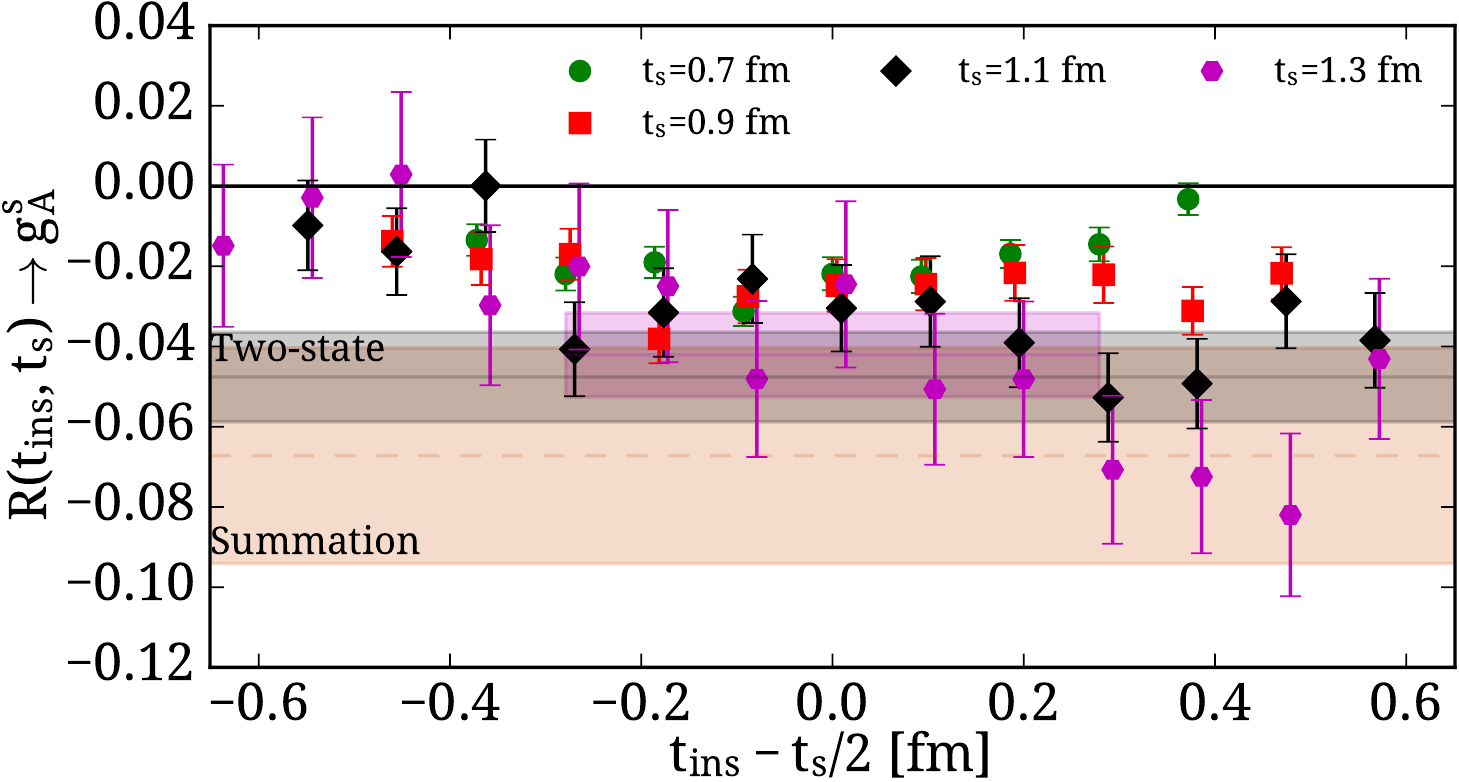}
        \includegraphics[width=\linewidth,angle=0]{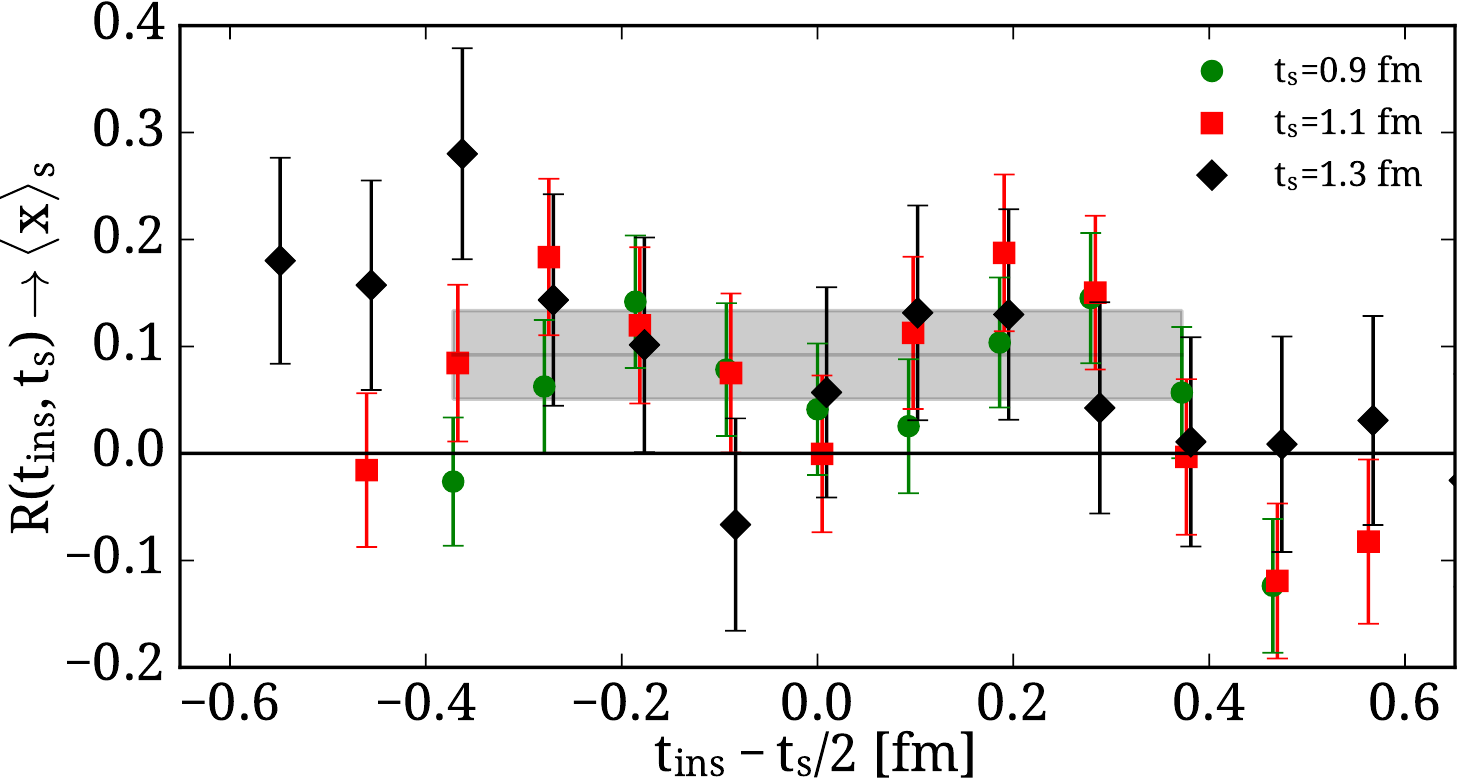}
      \end{minipage}
      \vspace{-0.5cm}
      \caption{\footnotesize Disconnected quantities for the strange quark mass. Here we drop deflation in favor of the TSM: the ultralocal employed
               63HP and 1024LP stochastic sources per configuration, whereas the one-derivative calculation used 30HP and 960LP sources. The number of
               configurations for both runs was increased to 2153, resulting in 861200 measurements.\label{plotStr}}
   \end{center}
\end{figure*}

\begin{figure*}[h!]
   \begin{center}
      \begin{minipage}{0.4\linewidth}
        \includegraphics[width=\linewidth,angle=0]{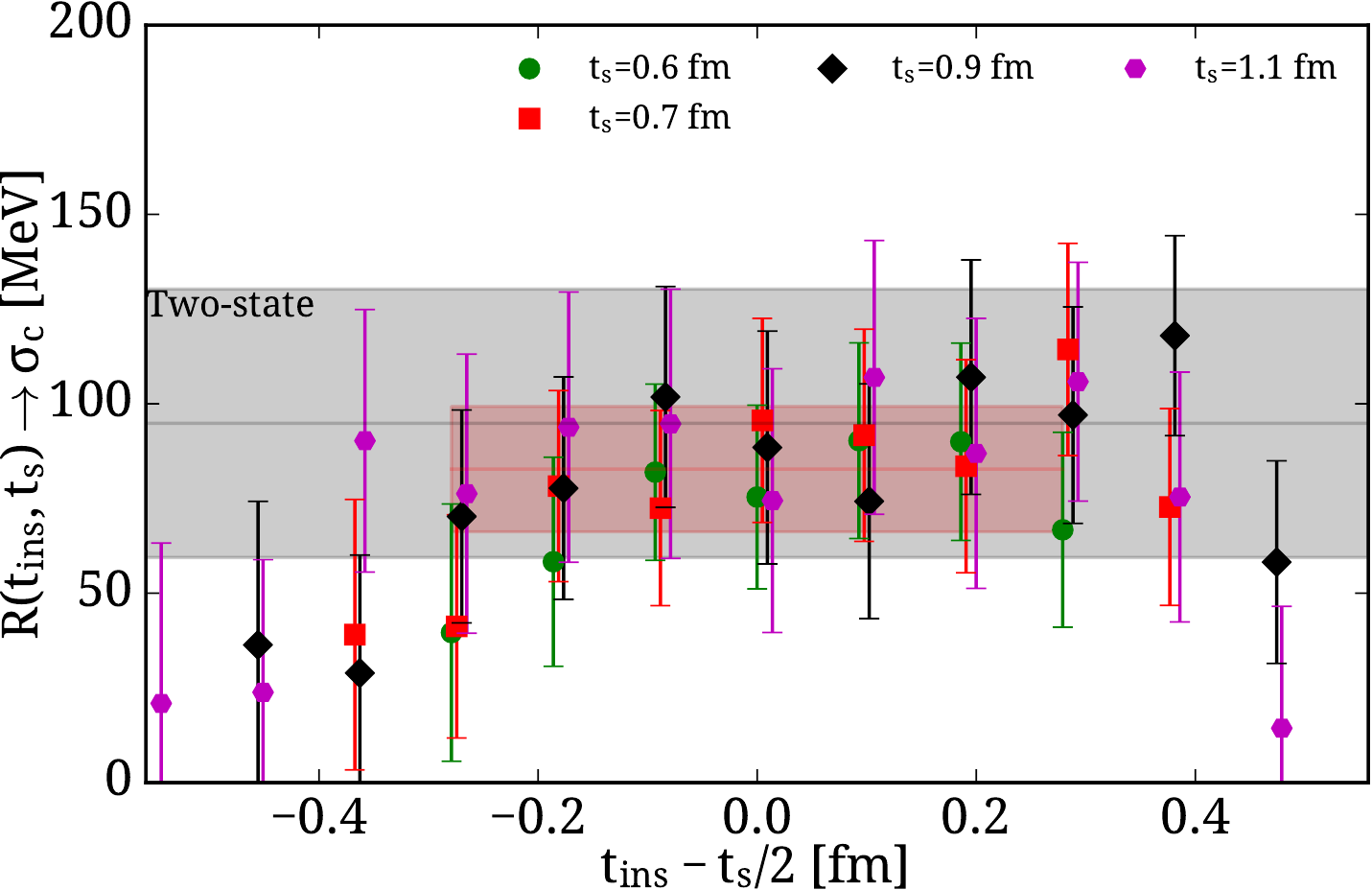}
      \end{minipage}
      \hspace{0.05\linewidth}
      \begin{minipage}{0.4\linewidth}
        \includegraphics[width=\linewidth,angle=0]{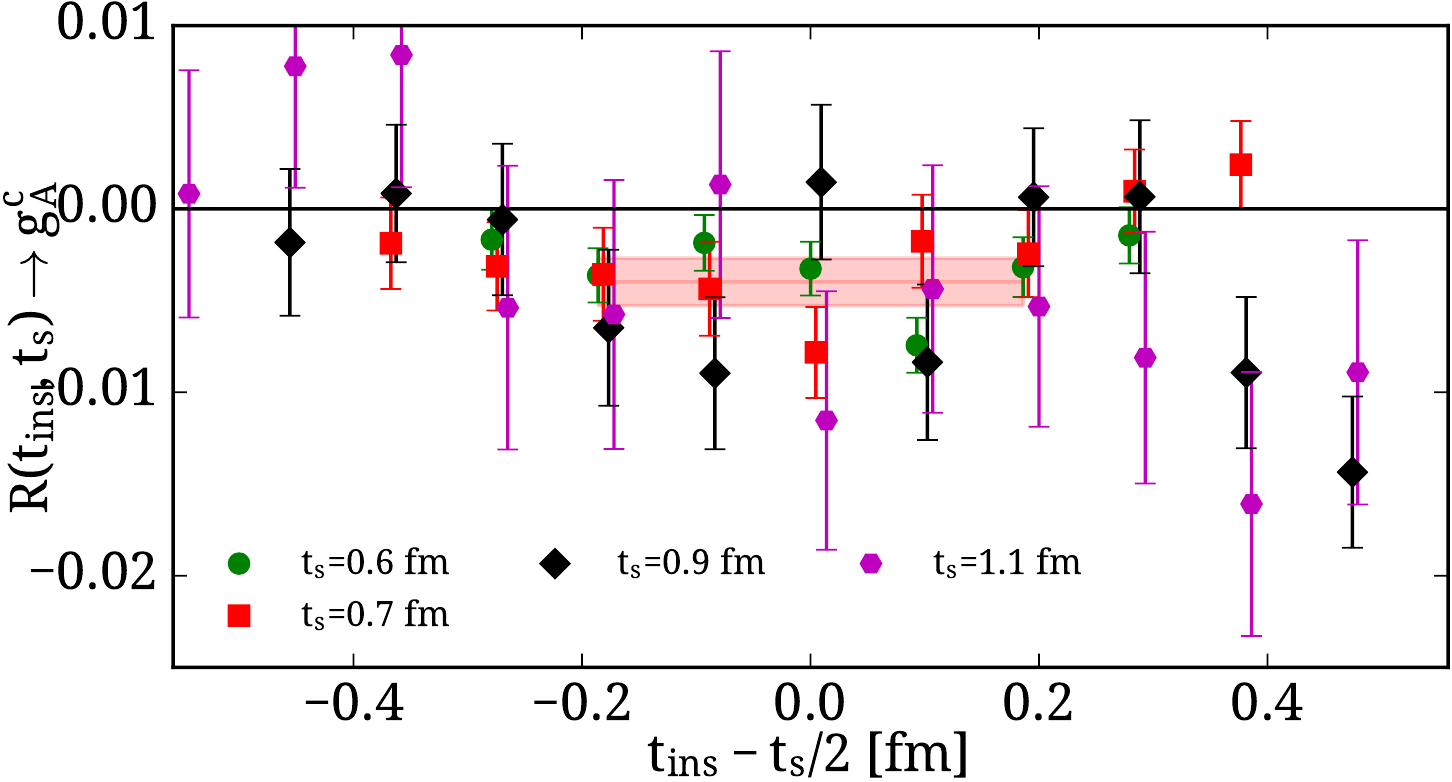}
        \includegraphics[width=\linewidth,angle=0]{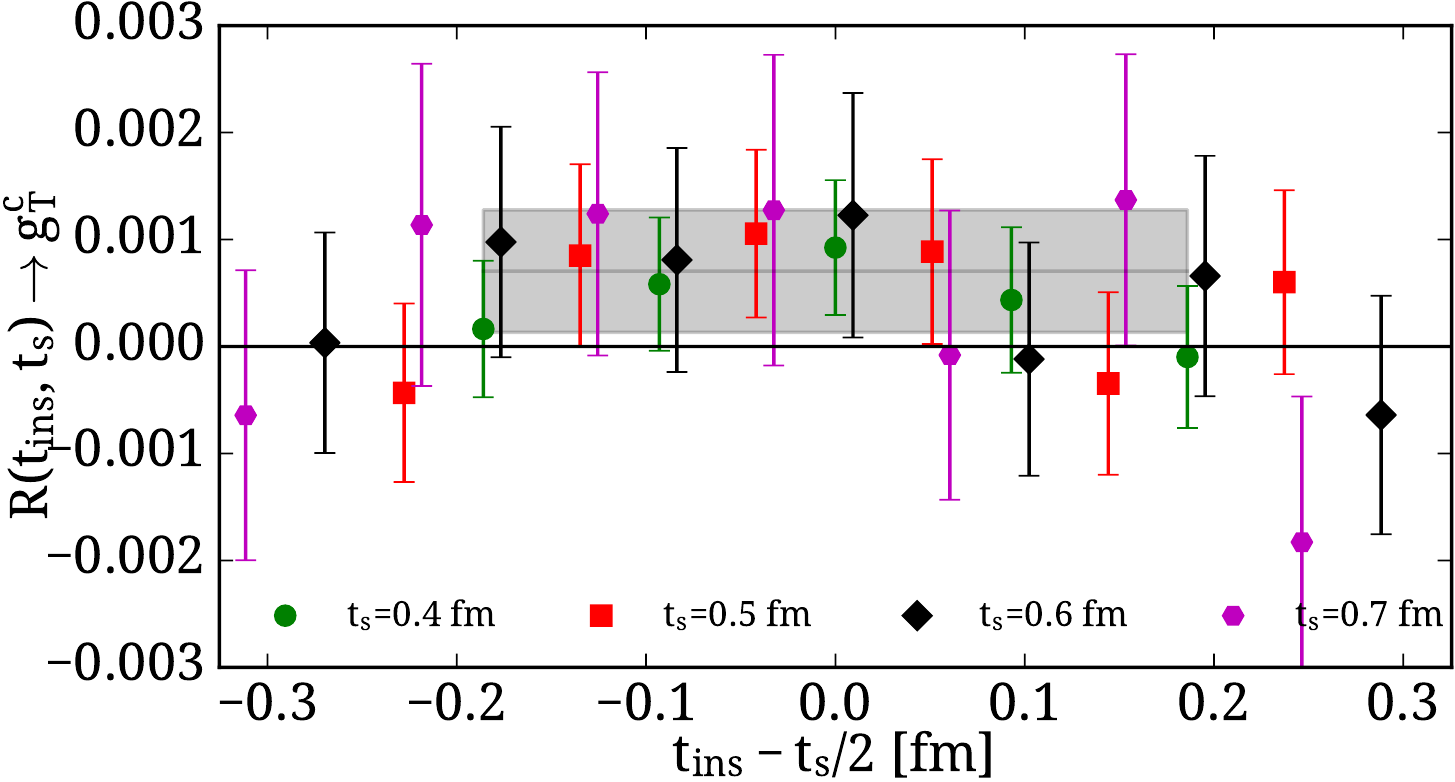}
      \end{minipage}
      \vspace{-0.5cm}
      \caption{\footnotesize Disconnected quantities for the charm quark mass. As in the strange case, we used solely the TSM as a variance reduction
               technique, and didn't try to invert faster with deflation. The charm used 5HP and 1250LP per configuration, and the total number of 
               configurations used was 2153, totaling 861200 measurements.\label{plotChr}}
   \end{center}
\end{figure*}

In general, the $\sigma-$terms and the axial charges give good signal for all the three masses; in particular, the upper plots of fig.~\ref{plotLgt} and
~\ref{plotStr} show a clear signal and good convergence. The charm (fig.~\ref{plotChr}) is noisier, but the absence of contributions of excited states
allows us to settle at a much lower value for the sink.

The tensor charge and the momentum fraction are more difficult to compute: for the tensor our signal is compatible with zero within $2\sigma$ for all the
masses, whewreas the momentum fraction gives a clear non-zero signal with large errors.

\section{Discussion}
\label{sec:con}
This work summarizes the effort of the EMTC to calculate disconnected diagrams directly at the physical pion mass, showing results of a high precision
computation of the disconnected contribution to all the charges and matrix elements of ultralocal and one-derivative operators for the nucleon, apart
from the electromagnetic form factors. Most of our results are in good agreement to what other groups have computed, and have led us to draw interesting
conclusions on the properties of the nucleon \cite{DinaPoS}; the rest are new quantities that have never been computed before, thus establishing a new
benchmark.

Our next steps are to analyze another two ensembles at the physical pion mass, one $N_f=2$ with larger volume, so we can estimate volume effects, and
another one with $N_f=2+1+1$ in order to assess the errors coming from the removal of the strange and charm sea quarks. We are also currently testing
new techniques that should allow us to compute the disconnected electromagnetic form factors with a reasonable stochastic error.

\acknowledgments
This work was supported, in part, by a grant from the Swiss National Supercomputing Centre (CSCS) under project ID s625. Additional computational resources
were provided by the Cy-Tera machine at The Cyprus Institute funded by the Cyprus Research Promotion Foundation (RPF),
${\rm NEAY\Pi O \Delta OMH}$/${\Sigma}$TPATH/0308/31.

\end{document}